# Energetic Particle and Other Space Weather Events of Solar Cycle 24

Nat Gopalswamy

*NASA Goddard Space Flight Center, Code 671, Greenbelt, MD 20771, USA*

**Abstract.** We report on the space weather events of solar cycle 24 in comparison with those during a similar epoch in cycle 23. We find major differences in all space weather events: solar energetic particles, geomagnetic storms, and interplanetary shocks. Dearth of ground level enhancement (GLE) events and major geomagnetic storms during cycle 24 clearly standout. The space weather events seem to reflect the less frequent solar eruptions and the overall weakness of solar cycle 24.

## INTRODUCTION

Large solar energetic particle (SEP) events and major geomagnetic storms are directly attributable to energetic coronal mass ejections (CMEs) from the Sun that have significant space weather implications [1,2]. SEP events with intensity ≥10 pfu in the >10 MeV energy channel are significant in causing space weather effects and are commonly referred to as large SEP events (1 pfu = 1 particle per $cm^2$.s.sr). Major geomagnetic storms are those with Dst index < -100 nT and are mostly caused by high-energy CMEs heading toward Earth. The most energetic CMEs originate only from solar active regions that can store and release large amounts of energy. Since large active regions appear during solar maxima, it is natural that most intense space weather events occur frequently during solar maxima. Therefore, considering various phases of the solar cycle (rise, maximum, and declining) is also important. Solar cycle 23 was the first cycle with near-complete observations of CMEs and their space weather consequences. In this paper, we compare the properties of cycle 24 space weather events (SEP events, geomagnetic storms, interplanetary shocks, and type II radio bursts) with those during the corresponding epoch of solar cycle 23. These observations have some important clues to the subdued nature of solar cycle 24.

## SEP EVENTS OF CYCLE 24

Table 1 lists the SEP events of cycle 24 (until the end of July 2012) with the proton intensity (Ip in pfu), solar source location, soft X-ray flare importance, the source active region number, CME onset time, CME speed (V km/s), and CME width (W in degrees; H=halo CME). The change in polar microwave brightness temperature and the "rush-to-the-poles" phenomenon observed in the prominence eruption locations, indicate that the Sun is already in the maximum phase in the northern hemisphere and is just beginning in the southern hemisphere [3]. The situation is similar to cycle 23, which had a polarity reversal in the north pole towards the end of year 2000 after ~4.5 years from the beginning of that cycle in May 1996 [4]. Therefore, it is appropriate to compare the cycle 24 activity up to the present time (end of July 2012) with that during the first 4.5 years of cycle 23 (end of year 2000). There were 27 large SEP events during the first 4.5 years of cycle 23, of which 22 had CME observations [5] that we use for comparison.

**TABLE 1.** List of large SEP events from solar cycle 24.

| Date and UT | Ip pfu | Location | Flare Imp. | NOAA AR # | CME UT | CME V | CME W |
|---|---|---|---|---|---|---|---|
| 2010/08/14 12:30 | 14 | N17W52 | C4.4 | 11099 | 10:12 | 1205 | H |
| 2011/03/08 01:05 | 50 | N24W59 | M3.7 | 11164 | 20:00$^c$ | 2125 | H |
| 2011/03/21 19:50 | 14 | N23W129 | Back | ???? | 02:24 | 1341 | H |
| 2011/06/07 08:20 | 72 | S21W54 | M2.5 | 11226 | 06:49 | 1255 | H |
| 2011/08/04 06:35 | 96 | N16W30 | M9.3 | 1 1261 | 04:12 | 1315 | H |
| 2011/08/09 08:45 | 26 | N17W69 | X6.9 | 11263 | 08:12 | 1610 | H |
| 2011/09/23 22:55 | 35 | N11E74 | X1.4 | 11302 | 10:48$^c$ | 1905 | H |
| 2011/11/26 11:25 | 80 | N27W49 | C1.2 | 11353 | 07:12 | 933 | H |
| 2012/01/23 05:30 | 3000$^a$ | N28W36 | M8.7 | 11402 | 04:12 | 2102 | H |
| 2012/01/27 19:05 | 800 | N27W71 | X1.7 | 11402 | 18:27 | 2408 | H |
| 2012/03/07 05:10 | 1500$^b$ | N17E27 | X5.4 | 11429 | 01:36 | 2544 | H |
| 2012/03/13 18:10 | 500 | N19W59 | M7.9 | 11429 | 17:36 | 1898 | H |
| 2012/05/17 02:10 | 255 | N11W76 | M5.1 | 11476 | 01:47 | 1618 | H |
| 2012/05/27 05:35 | 14 | N10W121 | Back | ???? | 20:57 | 2320 | H |
| 2012/06/16 19:55 | 14 | S17E06 | M1.9 | 11504 | 12:36$^d$ | 981 | H |
| 2012/07/07 04:00 | 25 | S18W51 | X1.1 | 11515 | 23:12$^c$ | 1912 | H |
| 2012/07/09 02:00 | 18 | S17W74 | M6.9 | 11515 | 16:54$^c$ | 1211 | H |
| 2012/07/12 18:55 | 96 | S15W01 | X1.4 | 11520 | 16:48 | 1360 | H |
| 2012/07/17 1715 | 136 | S28W65 | M1.7 | 11520 | 13:48 | 814 | 151 |
| 2012/07/19 08:00 | 70 | S28W75 | M7.7 | 11520 | 05:24 | 1681 | H |
| 2012/07/23 15:45 | 12 | S17W141 | Back | 11520 | 02:36 | 2142 | H |

$^a$shock peak: 6310 pfu; $^b$shock peak: 6300 pfu; $^c$Time corresponds to previous day; $^d$Time corresponds to two days before

There was no large SEP event for the first 2 years and seven months into cycle 24 and the first event occurred nearly 4 years after the last SEP event in cycle 23 (on 2006 December 13). The SEP events in 2010 and 2011 are all small, with the GOES proton intensity in the >10 MeV channel remaining below 100 pfu. SEP events with intensity exceeding 100 pfu occurred only in the maximum phase of cycle 24. By contrast, there were eleven >100 pfu events during the first 4.5 years of cycle 23, four of them occurring in the rise phase [5]. The source distributions on the solar disk and the flare sizes shown in Table 1 are similar to those observed during cycle 23. The average speed of the SEP-associated CMEs during cycle 24 is 1651 km/s (~20% higher) compared to 1373 km/s during the first 4.5 years of cycle 23 (see Fig. 1). The difference is more than the typical errors in the speed measurements. The average speed of all cycle-23 SEP-producing CMEs is 1621 km/s. There were only two CMEs in the cycle 23 SEP population with speed exceeding 2000 km/s (or 9%), but there are 6 such CMEs in cycle 24 (27%). Almost all the SEP-associated CMEs are full halos during cycle 24 (20 out of 21 or 95%), compared to only 68% (15 out of 22) during the corresponding epoch in cycle 23. The speed and width differences indicate that the cycle 24 CMEs were more energetic, yet they did not produce larger SEP events, suggesting that the cycle 24 CMEs are less efficient in accelerating energetic particles.

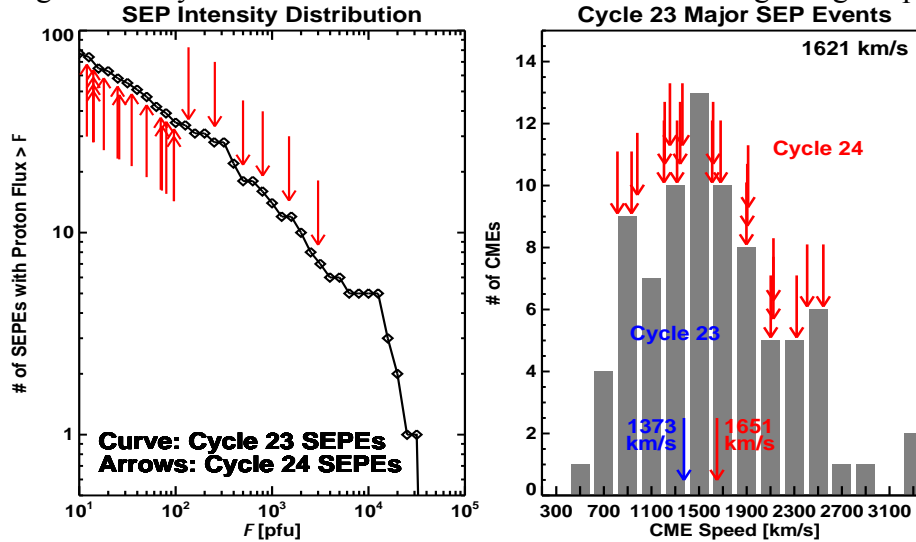

**FIGURE 1.** (left) Cumulative distribution of large SEP events (SEPEs) from cycle 23 with the sizes of the cycle 24 SEPEs marked. Down (up) pointing arrows indicate < 100 pfu (>100 pfu) events. All the > 100 pfu events are from the year 2012. In deciding event sizes, only early peak (before the shock peak) is considered. (right) The speed distribution of SEP-producing CMEs from solar cycle 23 with the speeds of cycle 24 events shown by arrows. The average speed of cycle 24 CMEs is 1651 km/s which is ~20% larger than the cycle-23 average speed (1373 km/s) over the first 4.5 years. The average speed of all SEP-producing CMEs of cycle 23 is 1621 km/s.

## SEP Butterfly Diagram Since 1976

Figure 2 shows the source latitudes of SEP events as a function of time since 1976 taken from the large proton events list maintained by NOAA's Space Weather Prediction Center: http://www.swpc.noaa.gov/ftpdir/indices/SPE.txt. We have included only those events that occurred on the front side of the Sun. Many SEP events originate from behind the west limb because the CMEs are generally very wide and the shocks driven by them have a larger extent, so they can quickly expand and cross the magnetic field lines connecting to an Earth observer. The approximate times of solar maximum of cycles 21-23 are marked in Fig.2. We call this plot as the SEP butterfly diagram, analogous to the sunspot butterfly diagram.

The SEP butterfly diagram shows that SEPs are essentially an active region phenomenon. We also note that the SEP events are generally clustered around the time of solar maximum, when large sunspot regions originate on the Sun that produce energetic CMEs. Generally one sees SEP sources in the northern and southern hemispheres with prominent asymmetries during some intervals. As noted before, cycle 24 showed severe north south asymmetry in that all but one of the fifteen large SEP events occurred in the northern hemisphere during the first 4.5 years. In this respect, cycle 24 resembles the rise and maximum phases of cycle 21: most of the events occurred in the northern hemisphere until after the maximum phase. Only in June 2012 SEP events have started appearing in the southern hemisphere.

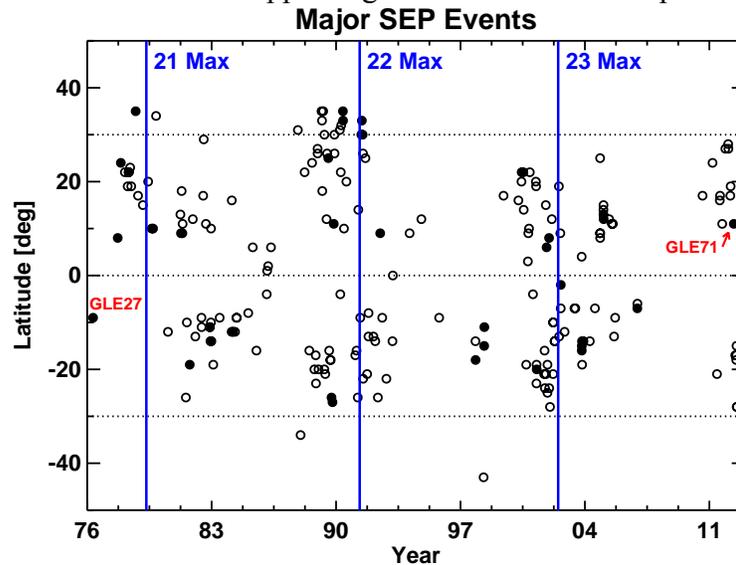

**FIGURE 2.** SEP butterfly diagram showing the solar sources of large SEP events observed by the GOES satellites since 1976. SEP events with ground level enhancement (GLE) are distinguished by filled circles. The first (#27) and last (#71) GLE events within the time span of this plot are noted. SEP events from backside eruptions and those occurring during data gaps have been excluded.

Figure 2 also shows the distribution of ground level enhancement (GLE) events (filled circles), that have GeV particles reaching Earth's atmosphere. It is remarkable that there was only one GLE event so far during cycle 24, whereas there were 5 such events in the first 4.5 years of cycle 23 (4 during the rise phase). The lone GLE event occurred from the northern hemisphere, where the maximum phase seems to coming to a close (see Fig. 3, updated from [3]). Figure 3 is the microwave butterfly diagram that shows the solar activity at low and high latitudes. The low-latitude emission (within $\pm30^\circ$) is due to sunspot activity representing the toroidal magnetic field of the Sun (gyro-resonance and free-free emission from active regions). The high-latitude emission (poleward of $60^\circ$ latitudes) is free-free emission from the polar chromosphere associated with enhanced polar magnetic field during solar minima. The current conditions in the north pole are similar to that around the year 2001 (cycle 23), while the maximum phase has started setting in in the south polar region.

One can see a clear association between the low-latitude magnetic activity and the occurrence of SEP events. During cycle 24, the sunspot activity appeared first in the northern hemisphere and was associated with most of the early SEP events. The activity in the southern hemisphere has a delay with respect to the northern hemispheric activity and only recently the activity picked up resulting in the last seven events in Table1. The separation between the SEP sources in the transition from cycle 22 to 23 (in the year 1996) is clearly much smaller than that from cycle 23 to 24 (in the year 2008), consistent with the prolonged minimum between cycles 23 and 24. This gap is even smaller (~0.5 years) between cycles 21 and 22 (see Fig. 2).

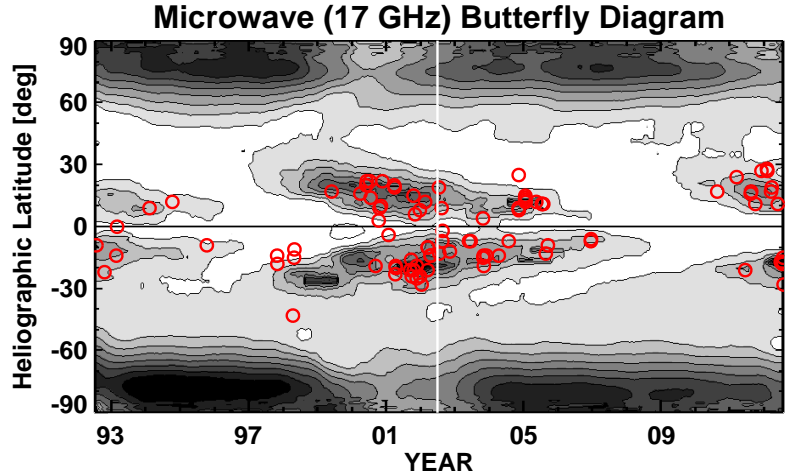

**FIGURE 3.** Microwave butterfly diagram showing the high and low-latitude activity of the Sun as observed by the Nobeyama radioheliograph at 17 GHz from July 1992 to the present (Carrington Rotation ending on June 21, 2012). The darker regions represent higher brightness temperature. The contour levels are at 10,000, 10,300, 10,609, 10,927, 11,255, 11,592, and 11,940 K. The vertical line marks the end of the solar maximum phase of solar cycle 23 (the time of completion of polarity reversal at solar poles). The solar maximum phase can be identified as the period of diminished polar microwave brightness at high latitudes and enhanced brightness at low latitudes. The microwave data corresponds to part of cycle 22, whole of cycle 23, and part of cycle 24. The circles denote the source location of the large SEP events over the period corresponding to the microwave observations.

## GEOMAGNETIC STORMS AND TYPE II BURSTS OF CYCLE 24

Table 2 lists the storm time, minimum Dst, CME onset, sky-plane speed V, X-ray flare onset, importance, location, and active region number for the 5 major (Dst < -100 nT) of cycle 24. In contrast, there were 35 such storms during the first 4.5 years of cycle 23 [6], clearly outnumbering cycle 24 with 7:1 margin. Furthermore, |Dst| ≤137 nT during cycle 24, compared to ≤301 nT for cycle 23 with 7 storms exceeding 200 nT. Examination of the solar wind plasma and magnetic signatures revealed that the southward component of the magnetic field causing the storms in Table 2 was in the sheath (2011 September and October), magnetic cloud (2012 March and July), and non-cloud ejecta (2012 April). No major CIR storm was observed in cycle 24 so far, compared to 3 during the first 4.5 years of cycle 23.

**TABLE 2.** Major geomagnetic storms of cycle 24 (Dst < -100 nT)

| Storm Date and Time | Dst (nT)[a] | CME Onset[b] | V km/s | Flare onset/Imp/Location/AR[c] |
|---|---|---|---|---|
| 2011/09/27 00:00 UT | -103 | 09/24 12:48 UT | 1915 | 12:33 M7.1 N15E58 1302 |
| 2011/10/25 02:00 UT | -137 | 10/22 01:25 UT | 593 | 01:20 C2.0 N40W30 DSF |
| 2012/03/09 09:00 UT | -133 | 03/07 01:36 UT | 2544 | 01:05 X1.3 N17E27 1429 |
| 2012/04/24 05:00 UT | -107 | 04/19 15:24UT | 400[d] | 13:42 --- S30E71 DSF |
| 2012/07/15 18:00 UT | -125 | 07/12 16:48 UT | 1360 | 15:37 X1.4 S15W01 1520 |

[a]From the World Data Center in Kyoto (http://wdc.kugi.kyoto-u.ac.jp/dst_realtime/index.html), [b]From http://cdaw.gsfc.nasa.gov and http://umbra.nascom.nasa.gov/lasco/observations/halo, [c]From http://www.swpc.noaa.gov/ftpdir/indices/events, [d]Preliminary, DSF = Disappearing Solar Filament

Table 3 compares the numbers of metric type II bursts (indicative of shocks near the Sun), Decameter-Hectometric (DH) type II bursts (shocks in the interplanetary medium), shocks detected in situ at Sun-Earth L1 by SOHO, and full halo CMEs between cycles 23 and 24 (considering only the first 4.5 years). All these are related phenomena, indicating overall low levels of activity during cycle 24. In particular, note the small number of major magnetic storms and the lone GLE event noted before. The magnetic content of CMEs reaching Earth seems to be low resulting in weak and infrequent geomagnetic activity. The weaker polar fields during cycle 23 seem to be reflected in the weaker sunspot activity during cycle 24.

**TABLE 3.** Comparison of shock-related phenomena during solar cycles 23 and 24.

| Event Type | #in Cycle 23[e] | # in Cycle 24 |
|---|---|---|
| Large SEP events (≥100 pfu) | 27 (11) | 21 (6) |
| Major magnetic storms (CME) | 35 | 5 |
| Major magnetic storms (CIR) | 3 | 0 |
| Metric Type II Bursts[a] | 310 | 130 |
| DH Type II Bursts[b] | 113 | 60 |
| Shocks at L1[c] | 92 | 59 |
| Full Halo CMEs[d] | 145 | 108 |

[a]ftp://ftp.ngdc.noaa.gov/STP/SOLAR_DATA/SOLAR_RADIO/SPECTRAL/Type_II;
[b]http://ssed.gsfc.nasa.gov/waves/data_products.html; [c]http://umtof.umd.edu/pm/FIGS.HTML;
[d]http://cdaw.gsfc.nasa.gov, [e]During the first 4.5 years

# SUMMARY AND CONCLUSIONS

We compared solar activity during solar cycles 23 and 24 and reported on the space weather events such as large SEP events and major geomagnetic storms. While the number of large SEP events of cycle 24 is similar to that in the corresponding epoch of cycle 23, the number of geomagnetic storms is very small. Overall energetic eruptions seem to be less frequent in cycle 24 as indicated by the lower number of type II radio bursts, full halo CMEs, and interplanetary shocks. The poor geoeffectiveness of cycle 24 CMEs and the extremely low frequency of GLE events (only one so far) are particularly noteworthy. Even though the CMEs associated with large SEP events are very energetic, they seem to be less efficient in accelerating particles. Further investigation is needed to understand the peculiarities of the weak solar cycle 24.

# ACKNOWLEDGMENTS

The author thanks P. Mäkelä and S. Yashiro for help with the figures. Work supported by NASA's LWS TR&T program.